\documentclass[12pt]{article}
\usepackage{amsmath}
\usepackage{amssymb}
\usepackage{amsfonts}

\numberwithin{equation}{section}

\title{
Rationally-extended Dunkl oscillator on the line}
\author{C Quesne\\ 
{\small Physique Nucl\'eaireTh\'eorique et Physique Math\'ematique,  Universit\'e Libre de Bruxelles,} \\ 
{\small Campus de la Plaine CP229, Boulevard~du Triomphe, B-1050 Brussels, Belgium}\\
{\small E-mail: Christiane.Quesne@ulb.be}}
\date{ }
\begin{document}
\baselineskip=22pt plus 1pt minus 1pt
\maketitle

\begin{abstract} 
It is shown that the extensions of exactly-solvable quantum mechanical problems connected with the replacement of ordinary derivatives by Dunkl ones and with that of classical orthogonal polynomials by exceptional orthogonal ones can be easily combined. For such a purpose, the example of the Dunkl oscillator on the line is considered and three different types of rationally-extended Dunkl oscillators are constructed. The corresponding wavefunctions are expressed in terms of exceptional orthogonal generalized Hermite polynomials, defined in terms of the three different types of $X_m$-Laguerre exceptional orthogonal polynomials. Furthermore, the extended Dunkl oscillator Hamiltonians are shown to be expressible in terms of some extended Dunkl derivatives and some anharmonic oscillator potentials.
\end{abstract}

\noindent
Keywords: quantum mechanics, Dunkl derivative, exceptional orthogonal polynomials 
%
%
\newpage
\section{Introduction}

Since several decades, deformations of quantum mechanics have been an intensive subject of research. In particular, the use of the reflection operator $R$, which is defined by $R f(x) = f(-x)$ and dates back to Wigner \cite{wigner} and Yang \cite{yang}, has found a lot of applications. More generally, Dunkl operators \cite{dunkl89}, which are sets of differential-difference operators associated with finite reflection groups, are very useful in several fields of mathematics and physics.\par
%
%
In mathematics, they play, for instance, an important role in the analysis of multivariate special functions and integral transforms associated with root systems \cite{dunkl91, dunkl08, dunkl14}. In physics, their usefulness for bosonizing supersymmetric quantum mechanics \cite{plyu96a, plyu96b} and generalizations thereof \cite{cq21}, describing anyons \cite{plyu94}, and constructing Dunkl supersymmetric orthogonal functions \cite{post, luo}, has been stressed. They are building blocks for describing an exchange operator formalism in Calogero-Moser-Sutherland models and their generalizations \cite{brink, poly, lapointe, cq95, khare}. They also provide a useful tool for proving the superintegrability of some models \cite{cq10a, cq10b}.\par
%
%
Several exactly-solvable quantum systems wherein the ordinary derivatives are replaced by Dunkl ones have been extensively studied. One may quote, for instance, the Dunkl oscillators in one, two, and three dimensions \cite{genest13a, genest14a, genest13b, genest14b, vinet}, the Dunkl-Coulomb problem in two and three dimensions \cite{genest15, ghazou19, ghazou20}, as well as the one-dimensional infinite \cite{chung19} and finite \cite{chung23} square wells. Coherent states \cite{ghazou22}, a generalization of shape invariance in supersymmetric quantum mechanics \cite{dong22}, Dunkl derivatives with two and three parameters \cite{chung21, dong21}, and some relativistic systems \cite{mota21, mota19} have also been investigated.\par
%
%
During the last few years, a lot of research activity has also been devoted to the construction of new exactly-solvable rational extensions of well-known quantum potentials. Such rational extensions arose after the introduction of exceptional orthogonal polynomials (EOPs) \cite{gomez09}, generalizing the classical orthogonal polynomials (COPs) and forming orthogonal and complete polynomial sets although they admit some gaps in the sequence of their degrees in contrast with the latter. $X_1$ EOPs were indeed shown to be related to the Darboux transformation in the context of shape invariant potentials in supersymmetric quantum mechanics \cite{cq08, cq09}. Infinite families of shape invariant potentials were then constructed in relation to $X_m$ EOPs \cite{odake09} and multi-indexed families of $X_{m_1 m_2 \ldots m_k}$ EOPs, connected with multi-step Darboux transformations, were considered \cite{gomez12, odake11}.\par
%
%
The purpose of the present paper is to combine the extensions of quantum mechanics connected with the use of Dunkl operators and that of EOPs. To show that this can be done, we plan to consider here the simplest Dunkl model, namely the one-dimensional Dunkl oscillator, and to employ its connection with the radial oscillator in order to construct some rationally-extended models. For such a purpose, we are going to use the three known infinite families of $X_m$-Laguerre EOPs related to one-step Darboux transformations \cite{cq09, grandati, cq11, liaw}.\par
%
%
This paper is organized as follows. In section~2, we review the treatment of the Dunkl oscillator on the line and its relation to the radial oscillator problem. In section~3, we summarize the results previously obtained for the three types of radial oscillator rational extensions. We apply the latter in section~4 to build rationally-extended Dunkl oscillators. In section~5, we show that the Hamiltonians of the latter can be expressed in terms of extended Dunkl operators. Finally, section~6 contains the conclusion.\par
%
%
\section{The one-dimensional Dunkl oscillator and its relation to the radial oscillator}

The one-dimensional Dunkl oscillator model is defined by the Hamiltonian \cite{genest13a}
\begin{equation}
  H_{\mu} = \frac{1}{2}(-D_{\mu}^2 + x^2) = \frac{1}{2}\left(- \frac{d^2}{dx^2} - \frac{2\mu}{x}\frac{d}{dx}
  + \frac{\mu}{x^2}(1-R) + x^2\right), \qquad -\infty<x<+\infty, 
\end{equation}
where $D_{\mu}$ is the Dunkl derivative
\begin{equation}
  D_{\mu} = \frac{d}{dx} + \frac{\mu}{x}(1-R), \qquad \mu > - \frac{1}{2},
\end{equation}
and $R$ is the reflection operator.\par
%
%
The corresponding Schr\"odinger equation reads
\begin{equation}
  H_{\mu} \psi^{(\mu)}_{2n+\epsilon}(x) = \left(2n+\epsilon+\mu+\frac{1}{2}\right) 
  \psi^{(\mu)}_{2n+\epsilon}(x), \qquad n=0, 1, 2, \ldots, \qquad \epsilon=0,1, \label{eq:HO}
\end{equation}
where the wavefunctions $\psi^{(\mu)}_{2n+\epsilon}(x)$, which are even for $\epsilon=0$ and odd for $\epsilon=1$, can be expressed as
\begin{equation}
  \psi^{(\mu)}_{2n+\epsilon}(x) = e^{-\frac{1}{2}x^2} H^{\mu}_{2n+\epsilon}(x)  \label{eq:wf} 
\end{equation}
in terms of generalized Hermite polynomials \cite{genest13a}. The latter are defined by
\begin{equation}
  H^{\mu}_{2n+\epsilon}(x) = (-1)^n \left(\frac{n!}{\Gamma(n+\mu+\epsilon+1/2)}\right)^{1/2} x^{\epsilon}
  L^{(\mu-1/2+\epsilon)}_n (x^2),  \label{eq:H}
\end{equation}
in terms of Laguerre polynomials \cite{chihara}. The wavefunctions (\ref{eq:wf}) satisfy the orthonormality property
\begin{equation}
  \int_{-\infty}^{+\infty} \left[\psi^{(\mu)}_{2n'+\epsilon'}(x)\right]^* \psi^{(\mu)}_{2n+\epsilon}(x)
  |x|^{2\mu} dx = \delta_{n',n} \delta_{\epsilon',\epsilon}.
\end{equation}
\par
%
%
These results for the Dunkl oscillator on the line can be directly obtained from well-known ones for the radial oscillator (see, e.g., ref.~\cite{cq09}). The corresponding Schr\"odinger equation reads
\begin{equation}
  \left(- \frac{d^2}{dx^2} + V_l(x)\right) \phi^{(l)}_n(x) = E^{(l)}_n \phi^{(l)}_n(x), \qquad n=0,1,2,\ldots.
  \label{eq:RO}
\end{equation}
Here
\begin{equation}
  V_l(x) = \frac{l(l+1)}{x^2} + \frac{1}{4} \omega^2 x^2, \qquad 0 < x < +\infty,
\end{equation}
where $l$ denotes the angular momentum quantum number and $\omega$ the frequency,
\begin{equation}
  E^{(l)}_n = \omega \left(2n+l+\frac{3}{2}\right),  \label{eq:HO-E}
\end{equation}
and
\begin{equation}
  \phi^{(l)}_n(x) = \left(\frac{\omega}{2}\right)^{\frac{1}{2}\left(l+\frac{3}{2}\right)} \left(\frac{2n!}
  {\Gamma(n+l+3/2)}\right)^{1/2} x^{l+1} L^{(l+1/2)}_n(z) e^{-\frac{1}{2}z}, \qquad z\equiv \frac{1}{2}
  \omega x^2,  \label{eq:RO-wf} 
\end{equation}
is such that
\begin{equation}
  \int_0^{+\infty} \left(\phi^{(l)}_{n'}(x)\right)^* \phi^{(l)}_n(x) dx = \delta_{n',n}.
\end{equation}
\par
%
%
On starting from the radial oscillator equation (\ref{eq:RO}), performing the replacements $\omega \to 2$, $l \to \mu -1$, and multiplying the wavefunction by $x^{-\mu}$, we indeed arrive at the equation
\begin{equation}
  \frac{1}{2} \left(- \frac{d^2}{dx^2} - \frac{2\mu}{x} \frac{d}{dx} + x^2\right) x^{-\mu} \phi^{(\mu-1)}_n(x)
  = \left(2n+\mu+\frac{1}{2}\right) x^{-\mu} \phi^{(\mu-1)}_n(x),
\end{equation}
agreeing with (\ref{eq:HO}) for $\epsilon=0$. Similarly, for $\omega \to 2$, $l \to \mu$, we get
\begin{equation}
  \frac{1}{2} \left(- \frac{d^2}{dx^2} - \frac{2\mu}{x}\frac{d}{dx} + \frac{2\mu}{x^2} + x^2\right) x^{-\mu}
  \phi^{(\mu)}_n(x) = \left(2n+\mu+\frac{3}{2}\right) x^{-\mu} \phi^{(\mu)}_n(x),
\end{equation}
giving back (\ref{eq:HO}) for $\epsilon=1$. In the transformation, equation (\ref{eq:RO-wf}) leads to equations (\ref{eq:wf}) and (\ref{eq:H}) after multiplying the former by an extra $1/\sqrt{2}$ factor, which takes the change of normalization from the half-line to the whole line into account (the phase factor in (\ref{eq:H}) being optional).\par
%
%
\section{Rational extensions of the radial oscillator}

\setcounter{equation}{0}

The rationally-extended radial oscillators resulting from one-step Darboux transformations read \cite{cq09, grandati, cq11}
\begin{align}
  V_{l, {\rm ext}}(x) &= V_l(x) + V_{l, {\rm rat}}(x) \nonumber \\
  &= V_{l'}(x) - 2 \frac{d^2}{dx^2} \log \varphi_{l'm}(x) - C,  \label{eq:V-ext}
\end{align}
where according to the choice made for $l'$ and $\varphi_{l'm}(x)$, three different types of extension are obtained:
\begin{eqnarray}
  & {\rm I:}\quad &  l' = l-1, \quad \varphi_{l-1,m}(x) \propto x^l e^{\frac{1}{4}\omega x^2} L^{(l-1/2)}_m
         (-\omega x^2/2), \quad C = - \omega; \nonumber \\
  & {\rm II:}\quad &  l' = l+1, \quad \varphi_{l+1,m}(x) \propto x^{-l-1} e^{-\frac{1}{4}\omega x^2} L^{(-l-3/2)}
         _m(\omega x^2/2), \quad C= \omega, \nonumber\\
  & &  \quad m < l+3/2; \\
  & {\rm III:}\quad & l'=l+1, \quad \varphi_{l+1,m}(x) \propto x^{-l-1} e^{\frac{1}{4}\omega x^2} L^{(-l-3/2)}_m
        (-\omega x^2/2), \quad C = - \omega, \nonumber \\
  & & \quad m<l+3/2, \quad \text{$m$ even}. \nonumber
\end{eqnarray}
The rational term in (\ref{eq:V-ext}) can be written as
\begin{equation}
  V_{l,{\rm rat}}(x) = - 2\omega \left\{\frac{\dot{g}^{(\alpha)}_m}{g^{(\alpha)}_m} + 2z 
  \left[\frac{\ddot{g}^{(\alpha)}_m}{g^{(\alpha)}_m} - \left(\frac{\dot{g}^{(\alpha)}_m}
  {g^{(\alpha)}_m}\right)^2\right]\right\}, 
\end{equation}
where $z = \frac{1}{2}\omega x^2$, $\alpha = l+\frac{1}{2}$, a dot denotes a derivative with respect to $z$, and
\begin{equation}
  g^{(\alpha)}_m(z) = \begin{cases}
      L^{(\alpha-1)}_m(-z) & \text{for type I}, \\[0.2cm]
      L^{(-\alpha-1)}_m(z), \quad m < \alpha+1 & \text{for type II}, \\[0.2cm]
      L^{(-\alpha-1)}_m(-z), \quad m < \alpha+1, \quad \text{$m$ even} & \text{for type III}.  
  \end{cases} \label{eq:g}
\end{equation}
The Schr\"odinger equations obtained by replacing $V_l(x)$ by $V_{l,{\rm ext}}(x)$ in (\ref{eq:RO}) assume different forms.\par
%
%
{}For type I, one gets
\begin{equation}
  \left(-\frac{d^2}{dx^2} + V_{l,{\rm ext}}(x)\right)  \phi^{({\rm I},l)}_n(x) = E^{({\rm I},l)}_n 
  \phi^{({\rm I},l)}_n(x),
\end{equation}
where
\begin{equation}
  E^{({\rm I},l)}_n = \omega \left(2n-2m+l+\frac{3}{2}\right) \label{eq:HO-E-I}
\end{equation}
and
\begin{equation}
  \phi^{({\rm I},l)}_n(x) = {\cal N}^{({\rm I},l)}_n 
  \frac{x^{l+1} e^{-\frac{1}{4}\omega x^2}}{L^{(l-1/2)}_m(-\frac{1}{2}\omega x^2)} 
  L^{{\rm I}, l+1/2}_{m,n}\left(\tfrac{1}{2}\omega x^2\right).
 \end{equation}    
Here $n = m, m+1, m+2, \ldots$ and $L^{{\rm I}, l+1/2}_{m,n}\left(\tfrac{1}{2}\omega x^2\right)$ is an $n$th-degree  $X_m$-Laguerre EOP of type {\rm I}. Such EOPs are known to form an orthogonal and complete polynomial set although 0, 1, \ldots, $m-1$ degrees are missing \cite{liaw}. The wavefunctions $\phi^{({\rm I},l)}_n(x)$ satisfy the orthonormality properties
\begin{equation}
  \int_0^\infty dx \left(\phi^{({\rm I},l)}_{n'}(x)\right)^* \phi^{({\rm I},l)}_n(x) = \delta_{n',n},  \label{eq:ortho}
\end{equation}
with the choice
\begin{equation}
  {\cal N}^{({\rm I},l)}_n = \left(\frac{\omega}{2}\right)^{\frac{1}{2}\left(l+\frac{3}{2}\right)}
  \left(\frac{2(n-m)!}{(n+l+\frac{1}{2}) \Gamma(n+l+\frac{1}{2}-m)}\right)^{1/2}
\end{equation}
for the normalization factor.\par
%
%
{}For type II, one gets
\begin{equation}
  \left(-\frac{d^2}{dx^2} + V_{l,{\rm ext}}(x)\right)  \phi^{({\rm II},l)}_n(x) = E^{({\rm II},l)}_n 
  \phi^{({\rm II},l)}_n(x),
\end{equation}
where
\begin{equation}
  E^{({\rm II},l)}_n = \omega \left(2n-2m+l+\frac{3}{2}\right) \label{eq:HO-E-II}
\end{equation}
and
\begin{equation}
  \phi^{({\rm II},l)}_n(x) = {\cal N}^{({\rm II},l)}_n 
  \frac{x^{l+1} e^{-\frac{1}{4}\omega x^2}}{L^{(-l-3/2)}_m(\frac{1}{2}\omega x^2)} 
  L^{{\rm II}, l+1/2}_{m,n}\left(\tfrac{1}{2}\omega x^2\right),
 \end{equation}    
provided $l>m-\frac{3}{2}$. Here $n = m, m+1, m+2, \ldots$ and $L^{{\rm II}, l+1/2}_{m,n}\left(\tfrac{1}{2}\omega x^2\right)$ is an $n$th-degree  $X_m$-Laguerre EOP of type {\rm II}. Such EOPs have rather similar properties as those of type I with 0, 1, \ldots, $m-1$ degrees missing \cite{liaw}. The wavefunctions $\phi^{({\rm II},l)}_n(x)$ satisfy orthonormality properties similar to (\ref{eq:ortho}) provided
\begin{equation}
  {\cal N}^{({\rm II},l)}_n = \left(\frac{\omega}{2}\right)^{\frac{1}{2}\left(l+\frac{3}{2}\right)}
  \left(\frac{2(n-m)!}{(n+l+\frac{3}{2}-2m) \Gamma(n+l+\frac{5}{2}-m)}\right)^{1/2}.
\end{equation}
\par
%
%
{}From (\ref{eq:HO-E}), (\ref{eq:HO-E-I}), and (\ref{eq:HO-E-II}), we note that type I or II extended radial oscillators have exactly the same spectrum as that of the starting radial oscillator $V_l(x)$. This is not the case for type III extensions, for which one gets
\begin{equation}
  \left(-\frac{d^2}{dx^2} + V_{l,{\rm ext}}(x)\right)  \phi^{({\rm III},l)}_n(x) = E^{({\rm III},l)}_n 
  \phi^{({\rm III},l)}_n(x),
\end{equation}
where
\begin{equation}
  E^{({\rm III},l)}_n = \omega \left(2n-2m+l+\frac{3}{2}\right) 
\end{equation}
and
\begin{equation}
  \phi^{({\rm III},l)}_n(x) = {\cal N}^{({\rm III},l)}_n 
  \frac{x^{l+1} e^{-\frac{1}{4}\omega x^2}}{L^{(-l-3/2)}_m(-\frac{1}{2}\omega x^2)} 
  L^{{\rm III}, l+1/2}_{m,n}\left(\tfrac{1}{2}\omega x^2\right),
 \end{equation}    
provided $l>m-\frac{3}{2}$ and $m$ is even. Here $n = 0, m+1, m+2, \ldots$ and $L^{{\rm III}, l+1/2}_{m,n}\left(\tfrac{1}{2}\omega x^2\right)$ is an $n$th-degree  $X_m$-Laguerre EOP of type {\rm III}. For such EOPs, 1, 2, \ldots\ $m$ degrees are indeed missing although they again form an orthogonal and complete polynomial set \cite{liaw}. Orthonormality properties similar to (\ref{eq:ortho}) are satisfied with the choice
\begin{equation}
  {\cal N}^{({\rm III},l)}_n = \left(\frac{\omega}{2}\right)^{\frac{1}{2}\left(l+\frac{3}{2}\right)} \times
    \begin{cases}
       \left(\frac{2}{\Gamma(l+\frac{3}{2}-m)m!}\right)^{1/2} & \text{if $n=0$}, \\[0.2cm]
       \left(\frac{2(n-m-1)!}{n \Gamma(n+l+\frac{3}{2}-m)}\right)^{1/2} & \text{if $n=m+1, m+2, \ldots$}.
    \end{cases}\end{equation}
\par
%
%
\section{Rational extensions of the Dunkl oscillator}

\setcounter{equation}{0}

To get rational extensions of the Dunkl oscillator from those of the radial oscillator, we are going to proceed as shown in section~2 for the standard problem. In all cases, we replace $\omega$ by 2, so that $z = \frac{1}{2}\omega x^2$ becomes $z=x^2$. In the $\epsilon=1$ case, which is obtained by introducing an operator $\frac{1}{2}(1-R)$ in the Hamiltonian, $l$ is replaced by $\mu$, so that the rational part of the potential is defined in terms of $\alpha=\mu+\frac{1}{2}$, whereas in the $\epsilon=0$ case, obtained by introducing an operator $\frac{1}{2}(1+R)$, $l$ is replaced by $\mu-1$, so that $\alpha-1$ will appear instead of $\alpha$ in the rational part of the potential. As a result, the three possible extensions of $H_{\mu}$ are given by
\begin{align}
  H_{\mu, {\rm ext}} &= \frac{1}{2} \Biggl\{- D_{\mu}^2 + x^2  - 2 \Biggl[\frac{\dot{g}^{(\alpha-1)}_m}
       {g^{(\alpha-1)}_m} + 2z\Biggl(\frac{\ddot{g}^{(\alpha-1)}_m}{g^{(\alpha-1)}_m} - \biggl(
       \frac{\dot{g}^{(\alpha-1)}_m}{g^{(\alpha-1)}_m}\biggr)^2\Biggr)\Biggr](1+R) \nonumber \\  
  &\quad - 2 \Biggl[\frac{\dot{g}^{(\alpha)}_m}
       {g^{(\alpha)}_m} + 2z\Biggl(\frac{\ddot{g}^{(\alpha)}_m}{g^{(\alpha)}_m} - \biggl(
       \frac{\dot{g}^{(\alpha)}_m}{g^{(\alpha)}_m}\biggr)^2\Biggr)\Biggr](1-R)\Biggr\},  \label{eq:hamiltonian}
\end{align}
where $g^{(\alpha)}_m(z)$ is defined as in (\ref{eq:g}) with $\alpha=\mu+\frac{1}{2}$ and $z=x^2$.\par
%
%
The corresponding Schr\"odinger equations now read
\begin{equation}
  H_{\mu,{\rm ext}} \hat{\psi}^{(\mu)}_{2n+\epsilon}(x) = \left(2n-2m+\epsilon+\mu + \frac{1}{2}\right)
  \hat{\psi}^{(\mu)}_{2n+\epsilon}(x),
\end{equation}
where $n=m, m+1, m+2, \ldots$ for type I or II, or $n=0, m+1, m+2, \ldots$ for type III and in all cases $\epsilon=0,1$. The corresponding wavefunctions $\hat{\psi}^{(\mu)}_{2n+\epsilon}(x)$ result from those of the extended radial oscillator after multiplication by $x^{-\mu}$ and a change of normalization from the half-line to the whole line. They may therefore be written as
\begin{equation}
  \hat{\psi}^{(\mu)}_{2n+\epsilon}(x) = e^{-\frac{1}{2}x^2} \hat{H}^{\mu}_{2n+\epsilon}(x),
\end{equation}
where $\hat{H}^{(\mu)}_{2n+\epsilon}(x)$ are exceptional orthogonal generalized Hermite polynomials, which we define by
\begin{align}
  \hat{H}^{(\mu)}_{2n+\epsilon}(x) &= \left(\frac{(n-m)!}{(n+\mu+\epsilon-\frac{1}{2})\Gamma(n-m+\mu+
  \epsilon-\frac{1}{2})}\right)^{1/2} \frac{x^{\epsilon}}{L^{(\mu+\epsilon-3/2)}_m(-x^2)} \nonumber \\
  &\quad \times L^{{\rm I},\mu-\frac{1}{2}+\epsilon}_{m,n}(x^2), \qquad \text{if $n=m, m+1, m+2, \ldots$},
\end{align}
for type I,
\begin{align}
  \hat{H}^{(\mu)}_{2n+\epsilon}(x) &= \left(\frac{(n-m)!}{(n-2m+\mu+\epsilon+\frac{1}{2})
        \Gamma(n-m+\mu+\epsilon+\frac{3}{2})}\right)^{1/2} \frac{x^{\epsilon}}
        {L^{(-\mu-\epsilon-1/2)}_m(x^2)} \nonumber \\
  &\quad \times L^{{\rm II},\mu-\frac{1}{2}+\epsilon}_{m,n}(x^2), \qquad \text{if $n=m, m+1, m+2, \ldots$},
\end{align}
for type II, and
\begin{align}
  \hat{H}^{(\mu)}_{2n+\epsilon}(x) &= \frac{x^{\epsilon}}{L^{(-\mu-\epsilon-1/2)}_m(-x^2)} \nonumber \\
  &\quad\times
    \begin{cases}
       \left(\frac{1}{\Gamma(\mu-m+\epsilon+\frac{1}{2})m!}\right)^{1/2} & \text{if $n=0$}, \\
       \left(\frac{(n-m-1)!}{n \Gamma(n-m+\mu+\epsilon+\frac{1}{2})}\right)^{1/2} 
            L^{{\rm III},\mu-\frac{1}{2}+\epsilon}_{m,n}(x^2) & \\
            \qquad\text{if $n=m+1, m+2, \ldots$} &   
    \end{cases}
\end{align}
for type III (with $m$ even). Note that the $\mu$ values have to be restricted to $\mu>\frac{1}{2}$ for type I or $\mu>m-\frac{1}{2}$ for type II or III (in order to get $\mu-\frac{1}{2}+\epsilon>0$ or $\mu-\frac{1}{2}+\epsilon>m-1$, respectively).\par
%
%
\section{Rationally-extended Dunkl oscillators in terms of extended Dunkl derivatives}

\setcounter{equation}{0}

In the present section, we plan to show that the rationally-extended Dunkl oscillator Hamiltonian (\ref{eq:hamiltonian}) can be rewritten in a simpler way 
\begin{equation}
  H_{\mu,{\rm ext}} = \frac{1}{2} \left(\hat{D}_{\mu}^2 + x^2 + G(x)\right) \label{eq:hamiltonian-bis}
\end{equation}
in terms of an extended Dunkl derivative
\begin{equation}
  \hat{D}_{\mu} = D_{\mu} + F(x) R = \frac{d}{dx} + \frac{\mu}{x}(1-R) + F(x)R,  \label{eq:dunkl-bis}
\end{equation}
defined in terms of some odd function $F(x)$, while $G(x)$ in (\ref{eq:hamiltonian-bis}) is some even function. The two functions $F(x)$ and $G(x)$ will of course be different for the three types of extensions.\par
%
%
On inserting definition (\ref{eq:dunkl-bis}) in (\ref{eq:hamiltonian-bis}) and comparing the result with equation (\ref{eq:hamiltonian}), one finds that $F(x)$ must satisfy the first-order differential equation
\begin{equation}
  F'(x) = 2 \Biggl\{\frac{\dot{g}^{(\alpha-1)}_m}
  {g^{(\alpha-1)}_m} + 2z\Biggl[\frac{\ddot{g}^{(\alpha-1)}_m}{g^{(\alpha-1)}_m} - \Biggl(
  \frac{\dot{g}^{(\alpha-1)}_m}{g^{(\alpha-1)}_m}\Biggr)^2\Biggr]\Biggr\}
  - 2 \Biggl\{\frac{\dot{g}^{(\alpha)}_m}
  {g^{(\alpha)}_m} + 2z\Biggl[\frac{\ddot{g}^{(\alpha)}_m}{g^{(\alpha)}_m} - \Biggl(
  \frac{\dot{g}^{(\alpha)}_m}{g^{(\alpha)}_m}\Biggr)^2\Biggr]\Biggr\},  \label{eq:F}
\end{equation}
where a dash denotes a derivative with respect to $x$, while a dot stands for a derivative with respect to $z=x^2$. In terms of such a function, $G(x)$ will then be given by
\begin{align}
  G(x) &=  -2 \Biggl\{\frac{\dot{g}^{(\alpha-1)}_m}
    {g^{(\alpha-1)}_m} + 2z\Biggl[\frac{\ddot{g}^{(\alpha-1)}_m}{g^{(\alpha-1)}_m} - \Biggl(
    \frac{\dot{g}^{(\alpha-1)}_m}{g^{(\alpha-1)}_m}\Biggr)^2\Biggr]\Biggr\}
    - 2 \Biggl\{\frac{\dot{g}^{(\alpha)}_m}
    {g^{(\alpha)}_m} + 2z\Biggl[\frac{\ddot{g}^{(\alpha)}_m}{g^{(\alpha)}_m} - \Biggl(
    \frac{\dot{g}^{(\alpha)}_m}{g^{(\alpha)}_m}\Biggr)^2\Biggr]\Biggr\} \nonumber \\
  &\quad + \frac{2\mu}{x} F(x) - F^2(x).  \label{eq:G}  
\end{align}
\par
%
%
Solving equation (\ref{eq:F}) is straightforward and leads to
\begin{equation}
  F(x) = 2x\left(\frac{\dot{g}^{(\alpha-1)}_m}{g^{(\alpha-1)}_m}- \frac{\dot{g}^{(\alpha)}_m}{g^{(\alpha)}_m}
  \right).  \label{eq:F-bis}
\end{equation}
On inserting (\ref{eq:F-bis}) in (\ref{eq:G}), one gets
\begin{equation}
  G(x) = 4(\alpha-1) \frac{\dot{g}^{(\alpha-1)}_m}{g^{(\alpha-1)}_m} - 4\alpha \frac{\dot{g}^{(\alpha)}_m}
  {g^{(\alpha)}_m} - 4z\left(\frac{\ddot{g}^{(\alpha-1)}_m}{g^{(\alpha-1)}_m} + \frac{\ddot{g}^{(\alpha)}_m}
  {g^{(\alpha)}_m}\right) + 8z \frac{\dot{g}^{(\alpha-1)}_m}{g^{(\alpha-1)}_m} \frac{\dot{g}^{(\alpha)}_m}
  {g^{(\alpha)}_m}.  \label{eq:G-bis}
\end{equation}
%
%
To obtain more explicit results for $F(x)$ and $G(x)$, one has to use definition (\ref{eq:g}) of $g^{(\alpha)}_m(z)$ in terms of Laguerre polynomials, as well as some known properties of the latter, as summarized in the appendix. The results read
\begin{equation}
  F(x) = 
  \begin{cases}
      2x\left(\frac{g^{(\alpha)}_{m-1}}{g^{(\alpha-1)}_m} - \frac{g^{(\alpha+1)}_{m-1}}{g^{(\alpha)}_m}\right) 
           & \text{for type I}, \\[0.2cm]
      2x\left(-\frac{g^{(\alpha-2)}_{m-1}}{g^{(\alpha-1)}_m} + \frac{g^{(\alpha-1)}_{m-1}}{g^{(\alpha)}_m}\right) 
           & \text{for type II},  \\[0.2cm]
      2x\left(\frac{g^{(\alpha-2)}_{m-1}}{g^{(\alpha-1)}_m} - \frac{g^{(\alpha-1)}_{m-1}}{g^{(\alpha)}_m}\right) 
           & \text{for type III}, 
  \end{cases}. \label{eq:F-ter}
\end{equation}
and
\begin{equation}
  G(x) = 
  \begin{cases}
      -4(\alpha-2+m)\frac{g^{(\alpha-1)}_{m-1}}{g^{(\alpha-1)}_m} + 4(\alpha-1+m)\frac{g^{(\alpha)}_{m-1}}
           {g^{(\alpha)}_m} & \text{for type I}, \\[0.2cm]
      4(\alpha-m)\frac{g^{(\alpha-1)}_{m-1}}{g^{(\alpha-1)}_m} -4(\alpha+1-m) \frac{g^{(\alpha)}_{m-1}}
           {g^{(\alpha)}_m} & \text{for type II},  \\[0.2cm]
      -4(\alpha-m)\frac{g^{(\alpha-1)}_{m-1}}{g^{(\alpha-1)}_m} + 4(\alpha+1-m)\frac{g^{(\alpha)}_{m-1}}
           {g^{(\alpha)}_m} & \text{for type III}. 
  \end{cases} \label{eq:G-ter}
\end{equation}
\par
%
%
{}For $m=1$, for instance, one gets as usual the same results for types I and II, namely
\begin{align}
  F(x) &= \frac{4x}{2x^2+2\mu-1} - \frac{4x}{2x^2+2\mu+1}, \nonumber \\
  G(x) &= - \frac{4(2\mu-1)}{2x^2+2\mu-1} + \frac{4(2\mu+1)}{2x^2+2\mu+1},
\end{align}
with $\mu>\frac{1}{2}$. For $m=2$, the results read
\begin{align}
  F(x) &= \frac{8x(2x^2+2\mu+1)}{(2x^2+2\mu+1)^2-2(2\mu+1)} - \frac{8x(2x^2+2\mu+3)}{(2x^2+2\mu+3)^2
      - 2(2\mu+3)}, \nonumber \\
  G(x) &= - \frac{8(2\mu+1)(2x^2+2\mu-1)}{(2x^2+2\mu+1)^2-2(2\mu+1)} + \frac{8(2\mu+3)(2x^2+2\mu+1)}
      {(2x^2+2\mu+3)^2-2(2\mu+3)},
\end{align}
with $\mu>\frac{1}{2}$ for type I,

\begin{align}
  F(x) &= \frac{8x(2x^2+2\mu-3)}{(2x^2+2\mu-3)^2+2(2\mu-3)} - \frac{8x(2x^2+2\mu-1)}{(2x^2+2\mu-1)^2
      + 2(2\mu-1)}, \nonumber \\
  G(x) &= - \frac{8(2\mu-3)(2x^2+2\mu-1)}{(2x^2+2\mu-3)^2+2(2\mu-3)} + \frac{8(2\mu-1)(2x^2+2\mu+1)}
      {(2x^2+2\mu-1)^2+2(2\mu-1)},
\end{align}
with $\mu>\frac{3}{2}$ for type II, and
\begin{align}
  F(x) &= \frac{8x(2x^2-2\mu+3)}{(2x^2-2\mu+3)^2+2(2\mu-3)} - \frac{8x(2x^2-2\mu+1)}{(2x^2-2\mu+1)^2
      + 2(2\mu-1)}, \nonumber \\
  G(x) &= - \frac{8(2\mu-3)(2x^2-2\mu+1)}{(2x^2-2\mu+3)^2+2(2\mu-3)} + \frac{8(2\mu-1)(2x^2-2\mu-1)}
      {(2x^2-2\mu+1)^2+2(2\mu-1)},
\end{align}
with $\mu>\frac{3}{2}$ for type III.\par
%
%
It is worth observing that in the transformation, the harmonic oscillator potential $\frac{1}{2}x^2$ becomes an anharmonic one $\frac{1}{2}[x^2+G(x)]$, which still vanishes at $x=0$ and goes to $+\infty$ at $x\to\pm\infty$, because $G(0) = G(\pm\infty) = 0$.\par
%
%
\section{Conclusion}

In the present work, we have shown that it is possible to combine the extensions of exactly-solvable quantum mechanical problems connected with the replacement of the ordinary derivative by a Dunkl one and with that of COPs by EOPs. For such a purpose, we have explicitly constructed three different types of rationally-extended Dunkl oscillators on the line, whose wavefunctions can be expressed in terms of exceptional orthogonal generalized Hermite polynomials, defined in terms of the three different types of $X_m$-Laguerre EOPs.\par
%
%
It is rather obvious that the method described here could be used to generate rationally-extended Dunkl oscillators associated to multi-indexed $X_{m_1m_2\ldots m_k}$-Laguerre EOPs, as well as rationally-extended Dunkl oscillators in more than one dimension.\par
%
%
Considering similar constructions for some other quantum potentials and a possible use of supersymmetric techniques would also be interesting problems for future study.\par
%
%
\section*{Acknowledgment}

The author was supported by the Fonds de la Recherche Scientifique - FNRS under Grant Number 4.45.10.08.\par
%
%
\section*{Data availability statement}

No new data were created or analysed in this study.\par
%
%
\section*{Appendix. Proof of equations (\ref{eq:F-ter}) and (\ref{eq:G-ter})}

\renewcommand{\theequation}{A.\arabic{equation}}
\setcounter{equation}{0}

In the proof of equations (\ref{eq:F-ter}) and (\ref{eq:G-ter}), we shall use the following well-known properties of Laguerre polynomials
\begin{align}
  &\left(z\frac{d^2}{dz^2} + (\alpha+1-z)\frac{d}{dz} + n\right) L^{(\alpha)}_n(z) = 0, \label{eq:L1} \\
  &\frac{d}{dz} L^{(\alpha)}_n(z) = - L^{(\alpha+1)}_{n-1}(z), \label{eq:L2}\\
  &z\frac{d}{dz} L^{(\alpha)}_n(z) = n L^{(\alpha)}_n(z) - (n+\alpha) L^{(\alpha)}_{n-1}(z),  \label{eq:L3} \\
  &L^{(\alpha-1)}_n(z) = L^{(\alpha)}_n(z) - L^{(\alpha)}_{n-1}(z).  \label{eq:L4}
\end{align}
\par
%
%
Equation (\ref{eq:F-ter}) directly results from equations (\ref{eq:F-bis}) and (\ref{eq:L2}) for the specific forms of $g^{(\alpha)}_m(z)$ given in (\ref{eq:g}).\par
%
%
To prove equation (\ref{eq:G-ter}) by starting from equation (\ref{eq:G-bis}), one first eliminates the second-order derivatives with the help of equation (\ref{eq:L1}), thereby leading to
\begin{equation}
  G(x) = 
    \begin{cases}
        4(2\alpha-2+z) \frac{\dot{g}^{(\alpha-1)}_m}{g^{(\alpha-1)}_m} + 4z \frac{\dot{g}^{(\alpha)}_m}
            {g^{(\alpha)}_m} - 8m + 8z \frac{\dot{g}^{(\alpha-1)}_m}{g^{(\alpha-1)}_m} \frac{\dot{g}^{(\alpha)}_m}
            {g^{(\alpha)}_m} & \text{for type I}, \\[0.2cm]
        -4z \frac{\dot{g}^{(\alpha-1)}_m}{g^{(\alpha-1)}_m} - 4(2\alpha+z) \frac{\dot{g}^{(\alpha)}_m}
            {g^{(\alpha)}_m} + 8m + 8z \frac{\dot{g}^{(\alpha-1)}_m}{g^{(\alpha-1)}_m} 
            \frac{\dot{g}^{(\alpha)}_m} {g^{(\alpha)}_m} & \text{for type II},  \\[0.2cm]
         4z \frac{\dot{g}^{(\alpha-1)}_m}{g^{(\alpha-1)}_m} - 4(2\alpha-z) \frac{\dot{g}^{(\alpha)}_m}
            {g^{(\alpha)}_m} - 8m + 8z \frac{\dot{g}^{(\alpha-1)}_m}{g^{(\alpha-1)}_m} \frac{\dot{g}^{(\alpha)}_m}
            {g^{(\alpha)}_m} & \text{for type III}. 
    \end{cases}
\end{equation}
\par
%
%
On applying equations (\ref{eq:L2}), (\ref{eq:L3}), and (\ref{eq:L4}), one gets
\begin{align}
  &4(2\alpha-2+z) \frac{\dot{g}^{(\alpha-1)}_m}{g^{(\alpha-1)}_m} = 4(2\alpha-2) \frac{g^{(\alpha)}_{m-1}}
      {g^{(\alpha-1)}_m} + 4m - 4(m+\alpha-2) \frac{g^{(\alpha-1)}_{m-1}}{g^{(\alpha-1)}_m}, \nonumber \\
  &4z \frac{\dot{g}^{(\alpha)}_m}{g^{(\alpha)}_m} = 4m - 4(m+\alpha-1) \frac{g^{(\alpha)}_{m-1}}
      {g^{(\alpha)}_m}, \\
  &8z \frac{\dot{g}^{(\alpha-1)}_m}{g^{(\alpha-1)}_m} \frac{\dot{g}^{(\alpha)}_m}{g^{(\alpha)}_m} = 
      - 8(\alpha-1) \frac{g^{(\alpha)}_{m-1}}{g^{(\alpha-1)}_m} + 8(\alpha-1+m) \frac{g^{(\alpha)}_{m-1}}
      {g^{(\alpha)}_m},  \nonumber
\end{align}
in the type I case,
\begin{align}
  &-4z \frac{\dot{g}^{(\alpha-1)}_m}{g^{(\alpha-1)}_m} = - 4m + 4(m-\alpha) \frac{g^{(\alpha-1)}_{m-1}}
      {g^{(\alpha-1)}_m}, \nonumber \\
  &-4(2\alpha+z) \frac{\dot{g}^{(\alpha)}_m}{g^{(\alpha)}_m} = 8\alpha \frac{g^{(\alpha-1)}_{m-1}}
      {g^{(\alpha)}_m} - 4m + 4(m-\alpha-1) \frac{g^{(\alpha)}_{m-1}}{g^{(\alpha)}_m}, \\
  &8z \frac{\dot{g}^{(\alpha-1)}_m}{g^{(\alpha-1)}_m} \frac{\dot{g}^{(\alpha)}_m}{g^{(\alpha)}_m} = 
      - 8(m-\alpha) \frac{g^{(\alpha-1)}_{m-1}}{g^{(\alpha-1)}_m} - 8\alpha \frac{g^{(\alpha-1)}_{m-1}}
      {g^{(\alpha)}_m},  \nonumber
\end{align}
in the type II case, and
\begin{align}
  &4z \frac{\dot{g}^{(\alpha-1)}_m}{g^{(\alpha-1)}_m} = 4m - 4(m-\alpha) \frac{g^{(\alpha-1)}_{m-1}}
      {g^{(\alpha-1)}_m}, \nonumber \\
  &-4(2\alpha-z) \frac{\dot{g}^{(\alpha)}_m}{g^{(\alpha)}_m} = -8\alpha \frac{g^{(\alpha-1)}_{m-1}}
      {g^{(\alpha)}_m} + 4m - 4(m-\alpha-1) \frac{g^{(\alpha)}_{m-1}}{g^{(\alpha)}_m}, \\
  &8z \frac{\dot{g}^{(\alpha-1)}_m}{g^{(\alpha-1)}_m} \frac{\dot{g}^{(\alpha)}_m}{g^{(\alpha)}_m} = 
      8(m-\alpha) \frac{g^{(\alpha-1)}_{m-1}}{g^{(\alpha-1)}_m} + 8\alpha \frac{g^{(\alpha-1)}_{m-1}}
      {g^{(\alpha)}_m},  \nonumber
\end{align}
in the type III case. It only remains to sum all the terms to get the final result (\ref{eq:G-ter}).\par
%
%

\end{document}